\documentstyle[prl,aps,epsfig,floats,twocolumn]{revtex}

\begin{document}

\title{Quenched Narrow-Line Laser Cooling of $^{40}$Ca
to Near the Photon Recoil Limit }

\author{E.~A.~Curtis$^*$, C.~W.~Oates, and L.~Hollberg}
\address{Time and Frequency Division, National Institute of
         Standards and Technology, \\325 Broadway, Boulder,
         Colorado 80305}


\maketitle

\begin{abstract}
We present a cooling method that should be generally applicable to
atoms with narrow optical transitions.  This technique uses
velocity-selective pulses to drive atoms towards a zero-velocity
dark state and then quenches the excited state to increase the
cooling rate.  We demonstrate this technique of quenched
narrow-line cooling by reducing the 1-D temperature of a sample of
neutral $^{40}$Ca atoms.  We velocity select and cool with the
$^1$S$_0$(4s$^2$)$\rightarrow$$^3$P$_1$(4s4p) 657 nm
intercombination line and quench with the
$^3$P$_1$(4s4p)$\rightarrow$$^1$S$_0$(4s5s) intercombination line
at 553 nm, which increases the cooling rate eight-fold. Limited
only by available quenching laser power, we have transferred 18 \%
of the atoms from our initial 2 mK velocity distribution and
achieved temperatures as low as 4 $\mu$K, corresponding to a
v$_{rms}$ of 2.8 cm/s or 2 recoils at 657 nm. This cooling
technique, which is closely related to Raman cooling, can be
extended to three dimensions.
\end{abstract}

\vspace{1cm}

Following the great success of sub-Doppler laser cooling
techniques with alkali atoms, several groups are now striving to
achieve similar cooling results for the alkaline-earth atoms, with
applications ranging from Bose-Einstein condensation to atomic
interferometry and optical frequency standards.  Unfortunately,
the lack of magnetic and hyperfine sub-structure in the ground
states of the predominant alkaline-earth isotopes precludes the
usual sub-Doppler mechanisms such as polarization-gradient
cooling~\cite{Dalibard,Chu}, velocity-selective coherent
population trapping~\cite{Aspect}, or Raman
cooling~\cite{Kasevich}. Furthermore, the broad cooling
transitions for the alkaline-earth atoms have large Doppler
limits, and typically yield magneto-optic trap (MOT) temperatures
of more than 1 mK~\cite{Shimizu,Mg,PTBMOT}.  These complications
have necessitated the development of alternative cooling
strategies to reach the microkelvin temperatures ``routinely"
achieved with several alkali species. This paper presents a new
approach, termed ``quenched narrow-line laser cooling" (QNLC),
which uses a narrow optical transition to provide increased
velocity selectivity for laser cooling, and quenching of the upper
state of this transition to enhance the cooling efficiency.  To
demonstrate the potential of this technique, we have used repeated
velocity-selective stepwise excitation with two intercombination
lines of neutral $^{40}$Ca to transfer as much as 18 \% of the
atoms in an initial velocity distribution with a 2 mK temperature
into a narrow peak. The lowest 1-D temperature observed was 4
$\mu$K with a corresponding transfer efficiency of 7 \%. Moreover,
these results can be significantly improved simply by increasing
the quenching power, and can be readily extended to two or three
dimensions. These greatly reduced temperatures should
significantly reduce the uncertainty in Ca-based optical frequency
standards, for which a recent measurement of the absolute
frequency was principally limited by residual atomic
velocity~\cite{Hg-Ca}.

Our approach to second-stage cooling takes advantage of the
two-electron structure of these atoms, which gives rise to narrow
intercombination lines. In principle, these transitions can serve
as excellent frequency or velocity (via the first-order Doppler
shift) discriminators, which are ideal for sub-recoil laser
cooling.  The spectacular results achieved with the
intercombination line of $^{88}$Sr, for which temperatures of 700
nK have been attained, confirm the potential of these
transitions~\cite{Katori,Kurt}.  Unfortunately, for several
species, such as Ca and Mg, the intercombination line is too weak
to effectively cool a distribution with an initial temperature of
several millikelvins. Nonetheless, Binnewies \textit{et
al.}~\cite{Maxwell} were able to use the velocity selectivity of
the Ca intercombination line at 657 nm to demonstrate a new
cooling mechanism, Maxwell-demon cooling. They were able to
achieve a net transfer of 5 \% of the initial distribution into a
peak with a 1-D temperature of $<$ 10 $\mu$K. However, this
technique does not seem to lend itself to 3-D cooling, which is
critical for many applications.

A more versatile approach is to increase the cooling efficiency by
quenching the upper level of the narrow transition through
excitation to another state that decays more quickly to the ground
state. In Fig. 1 we show how we quench the metastable $^3$P$_1$
state in Ca (lifetime = 400 $\mu$s) through excitation to the
$^1$S$_0$ (4s5s) level (lifetime = 30 ns) via the intercombination
line at 553 nm.  The enhancement of narrow-line cooling rates
through quenching was first demonstrated in the context of
sideband cooling of trapped ions~\cite{IonCool1,IonCool2}.  When
applied to neutral atoms, quenching can increase the cooling
efficiency not only by speeding up the rate at which one repeats
cooling cycles, but also by giving the atoms additional momentum
kicks towards zero velocity. These advantages were used to
increase the force on a metastable He atomic beam by excitation of
a two-photon transition with simultaneous light
fields~\cite{Wilbert}.

To take advantage of the velocity selectivity of the narrow
transition, however, it is essential to excite the atoms in a
stepwise manner with time-separated pulses. One can then implement
the velocity-selective excitation strategies that have been used
with great success in Raman cooling.  In the first demonstration
of Raman cooling, Kasevich and Chu used a Raman transition between
hyperfine ground states to cool Na atoms to a 1-D temperature of
100 nK and a 3-D temperature of 4.3 $\mu$K~\cite{Kasevich,Chu3D}.
Later Reichel \textit{et al.} used square Raman-cooling pulses to
achieve a 1-D temperature of 3 nK for Cs~\cite{French3D}.

For quenching times short compared to excitation times used for
the narrow transition, we propose the following QNLC procedure and
illustrate it with the Ca system.  We start by pumping a velocity
slice towards zero velocity using light whose frequency is tuned
slightly red of the $^1$S$_0$$\rightarrow$$^3$P$_1$
intercombination line at 657 nm (see Fig. 1). Due to the narrow
linewidth of this transition (400 Hz), the spectra of our red
pulses are Fourier-transform limited, so we can easily control the
width and shape of the velocity group excited. We then quench the
$^3$P$_1$ population with a 553 nm light pulse (copropagating with
the red pump beam), which moves the population to the
$^1$S$_0$(4s5s) state and gives the velocity-selected atoms a
second momentum kick towards zero velocity.  From this state the
atoms quickly return to the ground state via two cascaded decays
(see Fig. 1) with their associated (randomly directed) recoils.
Next we use a similar sequence of red and green pulses from the
opposite direction to pump a second velocity slice, symmetrically
located on the opposite side of the distribution, towards zero.  A
single cooling cycle can thus reduce atomic velocities on both
sides of the distribution by two-photon momenta (analogous to the
case of Raman cooling).  We repeat this cooling cycle many times,
driving the atoms towards zero velocity, which coincides with a
zero in our sinc$^2$ Fourier-transform-limited excitation
function. In this way, atoms that accumulate around zero velocity
have only a small probability of being pumped away, so in
principle, extremely low temperatures can be achieved.  As in the
case of Raman cooling, the cooling limit is set by the cooling
time and the width of the small transition-probability region
around zero velocity, but not by the random recoils.  Narrower
sinc$^2$ functions (i.e. longer red pulses in the time domain)
yield colder temperatures, but they address a smaller range of
velocities, so there is a compromise between temperature and
number, unless cooling with a sequence of pulse widths is
used~\cite{Kasevich,Chu3D,French3D}.

In our experimental demonstration of this technique, inadequate
light power at 553 nm (limited by what could be transferred
through a 180 m fiber) led to quenching times that were an order
of magnitude longer than the velocity-selective cooling pulses.
Under these conditions it was more efficient to use a slightly
modified version of the procedure just described, in which we use
a standing-wave pulse of 553 nm light to quench the $^3$P$_1$
state after each pair of counter-propagating red pulse
excitations.  The drawback to this approach is the randomness
introduced in the direction of the momentum kick from the
quenching pulse. This loss, however, is more than offset in our
case by the increased intensity of a standing (rather than
travelling) wave that reduces the quenching time and the number of
quenching pulses (by two) used per cooling cycle.

Our experimental realization of this cooling method (see Fig. 2)
begins with an $\sim$ 7 ms loading cycle, during which we load
10$^7$ atoms from a Ca beam into a MOT (see Ref.~\cite{NISTCaMOT}
for details of our diode-laser based MOT) using the 423 nm
$^1$S$_0$$\leftrightarrow$$^1$P$_1$ cooling transition (see Fig.
1). After this first cooling stage the temperature of the atomic
sample is about 2 mK (corresponding to a v$_{rms}$ of 0.6 m/s),
slightly above the 0.8 mK Doppler limit. To avoid large light
shifts of the ground state that would compromise the velocity
selectivity of the red cooling transition, we turn off the blue
trapping light for the duration of the second-stage cooling.  We
then commence QNLC with a pair of counter-propagating red pulses
(2.5 $\mu$s square-shaped $\pi$-pulses), separated by 2 $\mu$s in
time. This 657 nm light is spatially filtered with an optical
fiber and collimated to a diameter of 4 mm (8 mW in each beam). In
order to excite the m=0$\rightarrow$m=0 transition, the light is
linearly polarized parallel to the dominant B-field direction (due
to a trap imbalance, the atoms rest at a 6 G point in our magnetic
field gradient, which remains on during the entire measurement).
We tune the laser to a frequency $\sim$ 350 kHz (corresponding to
a velocity of 23 cm/s) red of resonance so that the first zero of
the sinc$^2$ frequency spectrum is within 5 kHz of resonance for
atoms at rest. The frequency of the red light is stabilized to an
environmentally isolated, high-finesse Fabry-Perot cavity
resonance, whose drift is cancelled to less than 0.3 kHz/minute
(typical data averaging times were about 1 minute).  After the
first pair of counter-propagating red pulses, atoms that were
transferred to the excited state are pumped to the $^1$S$_0$(4s5s)
m=0 state by a 553 nm pulse (duration of $\sim$ 50 $\mu$s) with an
efficiency of $\sim$ 50 \%. As a result of the small transition
rate (we estimate
$\Gamma$($^1$S$_0$(4s5s)$\rightarrow$$^3$P$_1$(4s4p)) $\sim$
2440(600) s$^{-1}$), we could only attain a 1/e quenching time of
55 $\mu$s, with 17 mW of green light collimated to a standing wave
diameter of 3 mm~\cite{PTBQuench}. The frequency-stabilized green
light is generated by a dye laser in another laboratory and sent
to us via optical fiber. This light intersects the atoms at an
angle of $\sim$ 8$^o$ relative to the red cooling beams.  From the
$^1$S$_0$(4s5s) state the atoms decay rapidly ($<$ 35 ns) to the
ground state by way of the $^1$P$_1$ state. Then follows a second
pair of counter-propagating red pulses, but this time in reverse
order (see Fig. 2) to make the overall cooling process more
symmetric. Due to the lobes of the sinc$^2$ frequency spectrum
there is some probability that the second pulse of a pulse pair
can transfer atoms excited by the first red pulse back to the
ground state, making the cooling due to the first pulse slightly
less efficient. After the second pair of red pulses, a second
green quenching pulse follows to complete the cooling cycle.  We
then repeat the cooling cycle (consisting of two pairs of
counter-propagating pulses, each followed by a standing-wave
quenching pulse) 8 to 20 times.  At the end of this sequence we
clean out the $^3$P$_1$ state with an extra green pulse lasting
100 to 150 $\mu$s that leaves less than 3 \% of the population in
the excited state.  We then measure the velocity distribution.

Our velocity probe consists of a single 657 nm pulse (10 or 20
$\mu$s in duration, depending on our desired spectral resolution)
that excites a narrow velocity slice of atoms to the excited
state.  It excites the m=0$\rightarrow$m=0 transition, and is
collinear with the red cooling beams.  While continuously cycling
the complete measurement sequence, we slowly sweep the frequency
of the velocity probe over the atomic velocity range to generate
our distributions.  We typically average 60 sweeps of 0.5 s
duration to generate a data set.  In order to achieve a good
signal-to-noise ratio for these velocity distribution
measurements, we use two 423 nm probe pulses in a normalized
shelving detection scheme that we developed for our optical clock
(see Fig. 2)~\cite{NISTCaMOT}.

In Figure 3 we show the effect of 15 cooling cycles taken with a
red pulse length of 2.5 $\mu$s and a green pulse length of 50
$\mu$s. We see that the QNLC process transfers most of the atoms
in the range of ±30 cm/s into the narrow peak at zero velocity,
consistent with the expected momentum transfer from our cooling
pulses.  A Gaussian fit to the central peak yielded a v$_{rms}$ of
6 cm/s, or about 4 recoils at 657 nm, corresponding to a 1-D
temperature of 17 $\mu$K.  The transfer efficiency can be
estimated by comparing the area of the peak to the area of the
initial distribution - in this case we find 30 \% transfer.
However, since approximately half of the atoms have escaped the
interaction region since we first turned off the trapping beams,
we have a net efficiency closer to 15 \% (we have seen net
efficiencies as high as 18 \%). Fast ballistic expansion due to
our warm initial temperature causes the atoms to move into weaker
parts of the laser beams during the cooling and probe periods.
Because of laser power constraints, we are unable to increase the
beam sizes, thus the number of cooling cycles we can use is
severely limited.

Indeed, when more cooling cycles are used, we see increased
transfer efficiency and narrower distributions (as was
demonstrated in Raman cooling
experiments~\cite{Kasevich,Chu3D,French3D}). However the longer
cooling time required means more atoms are lost transverse to the
cooling direction.  We experimented with using more cooling cycles
with less efficient quenching and found that the central peak was
fairly insensitive over the range from 18 cycles with 30 $\mu$s
quenching time to 8 cycles with 70 $\mu$s quenching time (keeping
the total cooling time approximately fixed at 1.5 ms).

The 14 cm/s width of the central peak in Fig. 3 is a result of the
convolution of the velocity distribution and the velocity probe.
Data taken at higher probe resolution indicate that v$_{rms}$ for
these conditions is less than 4 cm/s (7.5 $\mu$K). To strive for
colder temperatures, we increased the red pulse length to 5 $\mu$s
to give a sharper velocity discriminator.  Figure 4 shows a
velocity distribution resulting from 10 cooling cycles using these
longer cooling pulses.  To keep the 657 nm pulse area constant for
these measurements, we chose to double the size of the red beams
rather than reduce their power in an effort to increase their
pumping efficiency.  We also adjusted the detuning to $\sim$ -173
kHz (corresponding to 11.4 cm/s) to place the zero of the
excitation function at zero velocity.  Indeed, we see narrower
distributions under these conditions (we estimate a deconvolved
value of v$_{rms}$ $<$ 2.8 cm/s, corresponding to temperature of
3.7 $\mu$K) and the expected smaller fraction of atoms in the peak
(7 \% net efficiency) due to the reduced range of velocities
covered. We also see quite clearly several other nodes of the
sinc$^2$ function that serve as additional dark velocities where
the atoms can accumulate.  We fully expect that at higher
resolutions we can achieve sub-recoil temperatures, as was seen in
the Raman cooling experiments.  Monte Carlo simulations of the
cooling process (including the recoils at 553 nm, 1.03 $\mu$m, and
423 nm) support this expectation.

From these results, a strategy that can cool a large fraction of
the initial distribution to very cold temperatures becomes
evident, namely to use a series of pulses at different resolutions
and detunings~\cite{Kasevich,Chu3D,French3D}. To be able to
exploit such a strategy, however, we need to accelerate the
cooling process. With simple scaling of the 553 nm power, we
calculate that a fifteen-fold increase (dye lasers regularly
supply 20x the power we currently send to our atoms) would enable
us to cool the atoms nine times faster. Results from a Monte Carlo
simulation assuming a 1/e pumping time of 4 $\mu$s and using
multiple red cooling pulses ranging from 2.5 to 5 $\mu$s in
duration indicate that we can expect to transfer 50 \% of the
atoms into a peak with sub-recoil temperature.

Moreover, one can envision a feasible 3-D strategy in which one
cools alternately in three dimensions~\cite{Chu3D}, with each
reduction in temperature effectively increasing the available
cooling time. Due to the multi-recoils involved in this
experiment, it would be hard to achieve 3-D sub-recoil
temperatures due to recoil heating in transverse dimensions.
However, with appropriate shaping of our red
pulses~\cite{Chu3D,French3D} a limit of several recoils seems
quite feasible, especially in light of the near-recoil
temperatures achieved with 3-D Raman cooling~\cite{Chu3D}.
Furthermore, as one nears the recoil limit and presumably has
increased the cooling time available, one could remove the
quenching and use the decay of the $^3$P$_1$ state itself to
reduce the recoil effect. Regardless, 3-D temperatures of a few
microkelvins rather than millikelvins would greatly benefit the Ca
optical frequency standard, atomic interferometry, and other
applications.

In conclusion, we have proposed a second-stage cooling scheme
based on quenched narrow-line laser cooling.  We have used this
technique to reduce the temperature of a $^{40}$Ca atomic cloud in
1-D by a factor of 500. With ample quenching power, near
sub-recoil temperatures in 3-D should be possible. Moreover, this
technique should be applicable to other atoms (such as Mg) with
narrow optical transitions and available quenching transitions.

We thank J. Bergquist and U. Tanaka, who generously supplied the
553 nm light used in this work.

*Also at Phys. Dept., University of Colorado at Boulder, Boulder,
CO 80309.

\newpage

\begin{figure}[t]
\begin{center}
\includegraphics[width=8cm]{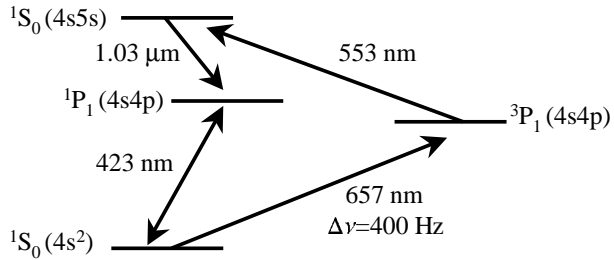}\vspace{1cm}
\vspace{0.3cm} \caption[]{Relevant energy level diagram for QNLC
in $^{40}$Ca. The 553 nm quenching beam transfers population from
the long-lived (400 $\mu$s) $^3$P$_1$(4s4p) state to the ground
state via the short-lived (30 ns) $^1$S$_0$(4s5s) state.
 } \label{Fig1}
\end{center}
\end{figure}


\begin{figure}[t]
\begin{center}
\includegraphics[width=8cm,angle=-180]{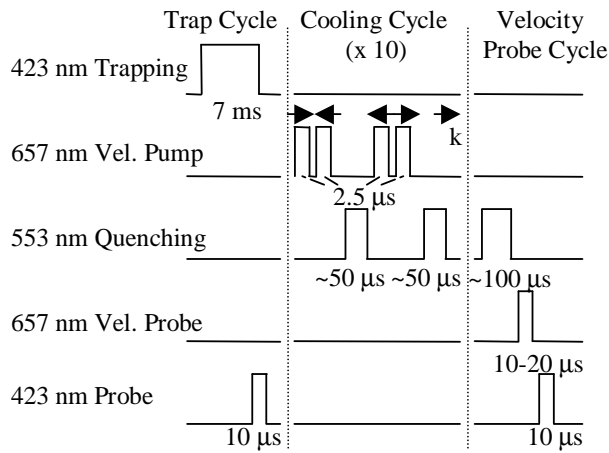}\vspace{1cm}
\caption[]{Timing diagram for our experimental realization of
QNLC.  Arrows represent the relative direction of the k-vector for
each pulse.  For short quenching times it would be more efficient
to use a co-propagating 553 nm pulse after each 657 nm pulse.}
\label{Fig2}
\end{center}
\end{figure}


\begin{figure}[t]
\begin{center}
\includegraphics[angle=-90,width=8cm]{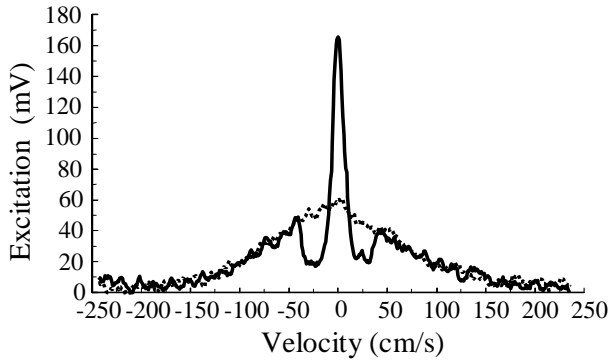}
\vspace{0.5cm} \caption[]{ Velocity distribution of $^{40}$Ca
atoms before (dotted line) and after (solid line) 15 second-stage
cooling cycles.  Each cycle consists of two 2.5 $\mu$s red pulse
pairs each followed by 50 $\mu$s of 553 nm quenching light.
Finally, a 553 nm light pulse (duration of 150 $\mu$s) is used to
pump remaining atoms back to the ground state.  The red velocity
probe pulse is 10 $\mu$s long.  We transfer about 15 \% of our
atoms into a narrow peak with a v$_{rms}$ of 5.9 cm/s,
corresponding to 4 recoils or a temperature of 17 $\mu$K.  This
width represents a convolution of the probe and the true velocity
distribution.} \label{Fig3}
\end{center}
\end{figure}


\begin{figure}[t]
\begin{center}
\includegraphics[angle=-90,width=8cm]{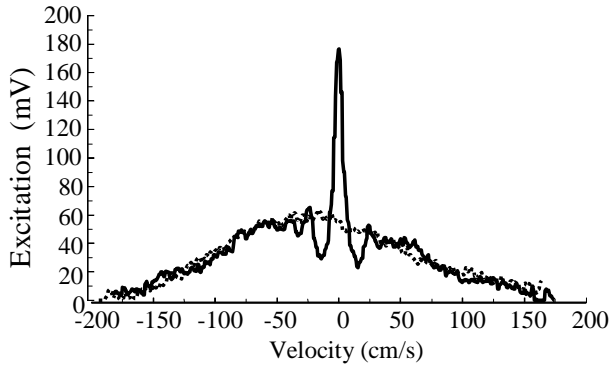}
\vspace{0.5cm} \caption[]{Velocity distribution as in Fig. 3
except with 10 second-stage cooling cycles with increased cooling
pulse duration (5 $\mu$s) and a higher resolution velocity probe
(20 $\mu$s duration).  A post-cooling pulse of 553 nm light
(duration of 100 $\mu$s) is used to pump remaining atoms back to
the ground state. The narrow peak has a v$^{rms}$ of 3.3 cm/s,
corresponding to $\sim$ 2 recoils or 5.3 $\mu$K. This width
represents a convolution of the probe with the actual velocity
distribution. } \label{Fig4}
\end{center}
\end{figure}

\end{document}